\documentclass[letterpaper]{article}

\PassOptionsToPackage{numbers,sort&compress,square}{natbib}
\usepackage[preprint]{neurips_2025}

\usepackage[utf8]{inputenc} % allow utf-8 input
\usepackage[T1]{fontenc}    % use 8-bit T1 fonts
\usepackage[hidelinks]{hyperref}       % hyperlinks
\usepackage{url}            % simple URL typesetting
\usepackage{booktabs}       % professional-quality tables
\usepackage{amsfonts}       % blackboard math symbols
\usepackage{nicefrac}       % compact symbols for 1/2, etc.
\usepackage{microtype}      % microtypography
\usepackage{xcolor}         % colors
\usepackage{orcidlink}
\usepackage{tabularx,booktabs,ragged2e,makecell,adjustbox,caption,placeins}
\newcolumntype{Z}{>{\RaggedRight\arraybackslash}X} % left-ish, wraps
\newcolumntype{Y}{>{\centering\arraybackslash}X}   % centered, wraps

\title{Data driven approaches in nanophotonics: A review of AI-enabled metadevices}

\author{%
  Huanshu Zhang \orcidlink{0009-0009-8332-7298}\\
  Department of Electrical Engineering\\
  The Pennsylvania State University\\
  University Park, PA 16802 \\
  \texttt{hpz5226@psu.edu} \\
  \And
  	Lei Kang \orcidlink{0000-0001-7718-7756}\\
  	Department of Electrical Engineering\\
  	The Pennsylvania State University\\
  	University Park, PA 16802 \\
  	\texttt{lzk12@psu.edu} \\
  	\And
  	Sawyer D. Campbell \orcidlink{0000-0002-3973-2730}\\
  	Department of Electrical Engineering\\
  	The Pennsylvania State University\\
  	University Park, PA 16802 \\
  	\texttt{sdc22@psu.edu} \\
  	\And
  	Jacob T. Young \orcidlink{0009-0001-3922-3768}\\
  	Department of Electrical Engineering\\
  	The Pennsylvania State University\\
  	University Park, PA 16802 \\
  	\And
  	Douglas H. Werner \orcidlink{0000-0001-5629-6478}\\
  	Department of Electrical Engineering\\
  	The Pennsylvania State University\\
  	University Park, PA 16802 \\
  	\texttt{dhw@psu.edu} \\
}

\makeatletter
\def\bstctlcite#1{\@bsphack
	\@for\@citeb:=#1\do{%
		\edef\@citeb{\expandafter\@firstofone\@citeb}%
		\if@filesw\immediate\write\@auxout{\string\citation{\@citeb}}\fi}%
	\@esphack}
\makeatother

\begin{document}
\bstctlcite{BSTcontrol}

\maketitle
% after \maketitle
\begingroup
\renewcommand\thefootnote{}\footnotetext{Accepted manuscript at \emph{Nanoscale}. DOI: \url{https://doi.org/10.1039/D5NR02043C}. The Version of Record may differ slightly.}
\addtocounter{footnote}{-1}
\endgroup

\begin{abstract}
Data-driven approaches have revolutionized the design and optimization of photonic metadevices by harnessing advanced artificial intelligence methodologies. This review takes a model-centric perspective that synthesizes emerging design strategies and delineates how traditional trial-and-error and computationally intensive electromagnetic simulations are being supplanted by deep learning frameworks that efficiently navigate expansive design spaces. We discuss artificial intelligence implementation in several metamaterial design aspects from high-degree-of-freedom design to large language model-assisted design. By addressing challenges such as transformer model implementation, fabrication limitations, and intricate mutual coupling effects, these AI-enabled strategies not only streamline the forward modeling process but also offer robust pathways for the realization of multifunctional and fabrication-friendly nanophotonic devices. This review further highlights emerging opportunities and persistent challenges, setting the stage for next-generation strategies in nanophotonic engineering. 
\end{abstract}

% --- Graphical TOC (alternate placement) ---
\begin{figure*}[t]
	\centering
	\includegraphics[width=\textwidth]{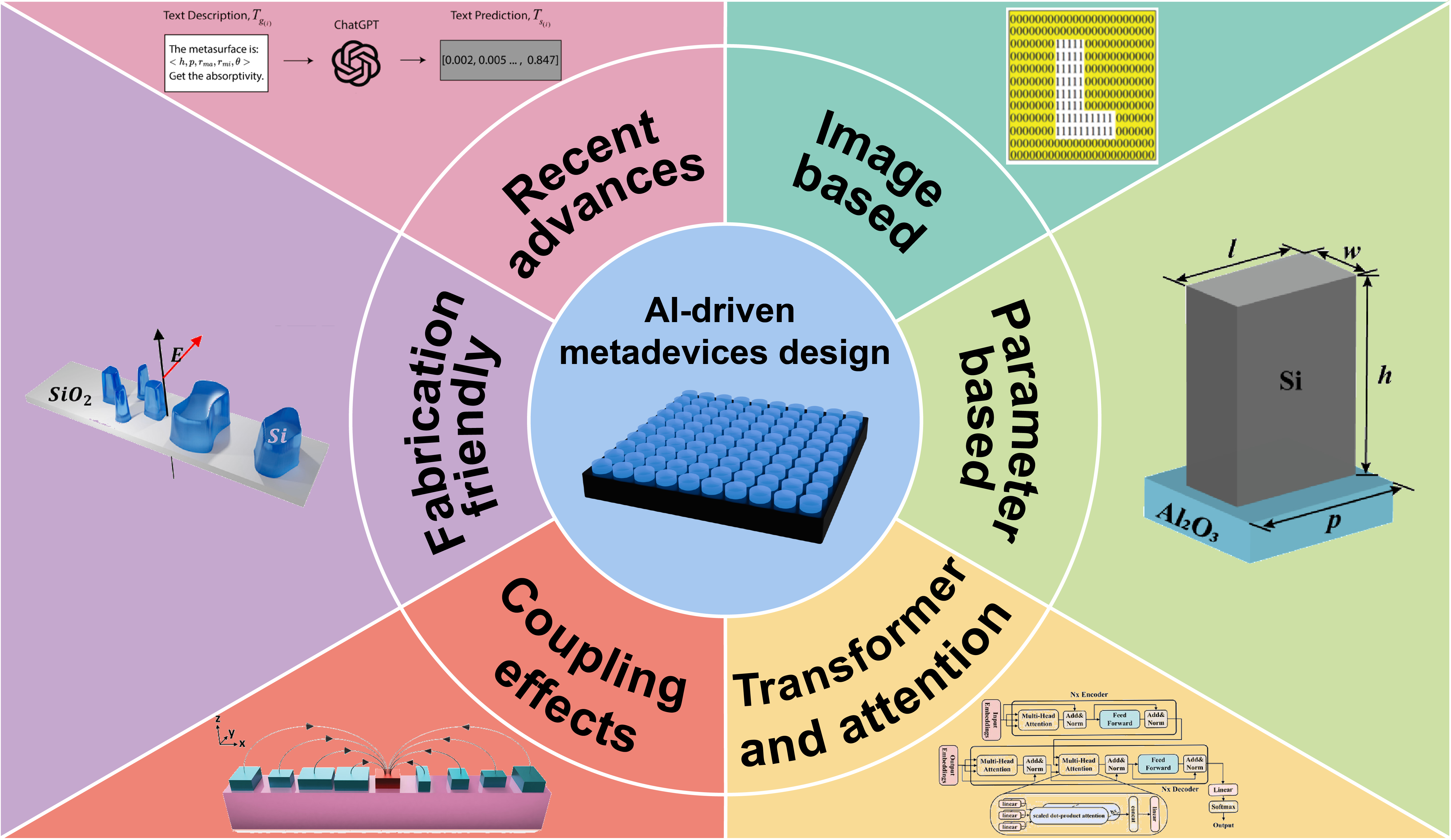}
	\caption*{Table of contents figure: AI surrogates are displacing brute-force simulations in meta-optics; we synthesize strategies from CNNs to LLMs that speed design, capture coupling, and yield fabrication-friendly, multifunctional devices.}
	\label{fig:toc}
\end{figure*}
% -------------------------------------------

\section{Introduction}
The emergence of artificial structures not only unveils the complexity of light--matter interaction, which generally occurs within atoms or molecules, but also provides an unprecedented opportunity for control of light via engineered devices. Compared with photonic crystals which rely on collective response arising from periodic perturbations, metamaterials have demonstrated powerful and versatile manipulation of electromagnetic (EM) waves via structural engineering of their subwavelength building blocks, i.e., meta-atoms~\citep{zheludev_metamaterials_2012}. For instance, metamaterials exhibiting properties that are not usually seen in naturally occurring materials, such as optical magnetism~\citep{cai_metamagnetics_2007}, negative effective refractive index~\citep{valentine_three-dimensional_2008}, and strong chirality~\citep{gansel_gold_2010} have been reported. Importantly, those exotic properties primarily arise from a meta-atoms’ architecture rather than the intrinsic properties of the base materials. These characteristics make metamaterials an excellent candidate for applications in devices that require precise control over the intrinsic properties (magnitude, phase, polarization, etc.) of light. Extending these principles to 2D has led to the development of metasurfaces which gain their properties from single-layer or few-layer planar artificial structures. In contrast to metamaterials, metasurfaces provide a more compact and fabrication-friendly approach to light control~\citep{chen_review_2016}, facilitating applications such as wavefront shaping~\citep{yu_light_2011}, beam steering~\citep{huang_dispersionless_2012}, holography~\citep{huang_metasurface_2018}, optical computing~\citep{zhou_optical_2024}, etc. The development of metadevices as an extension of the metamaterial and metasurface paradigm paves the way toward the next generation of photonic technologies.

Though metamaterials and metasurfaces have manifested an impressive capability to manipulate light, the design of metadevices, especially those for sophisticated and/or multiple functionalities can be an extremely challenging engineering task. Metamaterial design has traditionally relied on iterative numerical simulations that solve Maxwell’s equations to determine the optical/electromagnetic response of its constituent meta-atoms. Common computational tools include the Finite-Difference Time-Domain (FDTD) method~\citep{hao_fdtd_2009}, and Finite-Element Method (FEM)~\citep{jin_finite_2014}, as they discretize EM fields in space and/or time to analyze how nanostructures interact with EM waves. However, given the fact that meta-atoms generally include deep-subwavelength features, accurate metamaterial simulations based on these methods can be computationally expensive~\citep{dong_advanced_2025}, especially when considering multiscale devices. On the other hand, optimization methods such as genetic algorithms (GA) or other evolutionary optimization techniques are commonly employed to design metamaterials structures~\citep{campbell_review_2019}. The optimization iteratively refines designs by evaluating optical properties of the structure using numerical EM solvers (commercial or customized), updating the design based on an objective function, and repeating the process until some stopping criteria is met~\citep{elsawy_numerical_2020}. Despite previous efforts made to improve the efficiency of both optimization methods and EM solvers, these approaches still impose a significant computational burden in terms of time and resources. Consequently, evaluating and optimizing subwavelength architecture remains a great challenge, necessitating more advanced and efficient strategies.

Machine learning (ML) and deep learning (DL), subsets of artificial intelligence (AI)~\citep{russell_artificial_2021}, offer data-driven approaches to identifying complex structure/response correlations in metamaterial design due to their ability to replicate hyper-dimensional non-linear relationships and, once trained, near instantaneous evaluation. ML algorithms include supervised, unsupervised, and reinforcement learning, with supervised learning being the most utilized due to its structured data-label relationships~\citep{song_artificial_2024}. Fully connected (FC) neural networks, such as multilayer perceptrons (MLPs), have been used to map design parameters to optical properties~\citep{malkiel_deep_2017,tahersima_deep_2019,ma_deep-learning-enabled_2018,peurifoy_nanophotonic_2018}. Recurrent neural networks (RNNs), including long short-term memory (LSTM)~\citep{hochreiter_long_1997} networks, excel in sequential data, making them suitable for applications involving sequence processing such as in the case of spectra~\citep{deng_long_2022,pillai_leveraging_2021}. Convolutional neural networks (CNNs), designed for image processing, effectively analyze spatial patterns within metasurfaces~\citep{li_deep_2022,an_deep_2020}. The transformer architecture, first introduced by Vaswani \textit{et al.} in 2017~\citep{vaswani_attention_2017}, uses self-attention mechanisms to capture complex data dependencies, enabling parallelized training and specializing in sequence modelling, forming the basis for large language models (LLMs) like ChatGPT~\citep{openai_gpt-4_2023}, has also been applied in designing metamaterials~\citep{chen_broadband_2023}.

DL has rapidly advanced metamaterial modelling and optimization~\citep{ueno_ai_2024,khaireh-walieh_newcomers_2023}. The journey began in 2017 when Malkiel \textit{et al.} developed a bidirectional MLP-based DL model for designing H-shaped plasmonic nanostructures~\citep{malkiel_deep_2017}. Ma \textit{et al.} (2018) further advanced the method by developing two bidirectional neural networks with partial stacking for both forward and inverse modelling of reflection and CD spectra~\citep{ma_deep-learning-enabled_2018}, while Peurifoy \textit{et al.} (2018) applied MLPs to optimize multilayer nanoparticles~\citep{peurifoy_nanophotonic_2018}. Asano \textit{et al.} (2018) employed CNNs to enhance the Q-factor of photonic crystals~\citep{asano_optimization_2018}. Sajedian \textit{et al.} (2019) integrated CNNs and RNNs to predict absorption spectra of random plasmonic nanostructures~\citep{sajedian_finding_2019}. Chen \textit{et al.} (2023) recently harnessed a transformer-based model for designing broadband solar metamaterial absorbers~\citep{chen_broadband_2023}.

A few recent review papers on AI-assisted metamaterial design have charted diverse paths toward solving forward and inverse design challenges. For instance, Masson \textit{et al.}~\citep{masson_machine_2023} discussed the use of ML for nanoplasmonics, revealing how advanced algorithms uncover the complex structure–property relationships that underpin high-performance device engineering. Chen \textit{et al.}~\citep{chen_artificial_2022} bridged the fields of AI and meta-optics by detailing how AI accelerates both design and functional realization of flat optical devices. Furthermore, Ueno \textit{et al.}~\citep{ueno_ai_2024} offered a perspective that zeroes in on AI-enabled design-for-manufacturing and computational post-processing to help mitigate the simulation–fabrication gaps in metasurfaces. Other review articles~\citep{dong_advanced_2025,khaireh-walieh_newcomers_2023,fu_unleashing_2024,xu_interfacing_2021,jin_intelligent_2022,yao_intelligent_2019,hegde_deep_2020,wiecha_deep_2021,ma_deep_2021,piccinotti_artificial_2021,tezsezen_ai-based_2024,wang_advancing_2022,jiang_deep_2020,qian_progress_2025,campbell_explosion_2021} have been dedicated to different aspects such as free-form optimization~\citep{park_free-form_2022}, light–matter interactions~\citep{midtvedt_deep_2022}, physics-informed neural networks~\citep{ji_recent_2023}, and intelligent inverse design for phononic metamaterials~\citep{jin_intelligent_2022}. In contrast, from a model-centric perspective, this review will primarily focus on emerging design strategies. In particular, we aim to discuss high-degree-of-freedom (DoF) metamaterial design, the use of transformers and attention mechanisms, the prediction of mutual coupling effects between meta-atoms, strategies for designing robust and fabrication-friendly metamaterials, and recent advances as well as future prospects. We also summarize the computational costs in practice for those reported methods. Across the studies we survey, the dominant expense is usually dataset generation (e.g., full-wave simulations), while model training is a secondary, one-off cost and inference is typically near-interactive on commodity GPUs. For inverse design, simpler models such as one-shot / tandem / autoencoder-based approaches tend to yield faster per-candidate sampling than sequential samplers (e.g., diffusion), which trade speed for stability/diversity. Consequently, hardware requirements scale most strongly with training-set size, fidelity and the inverse-design strategy, rather than with the specific deep-learning library. In practice, a single consumer GPU is usually sufficient for training and inference, whereas multi-GPU or cluster access mainly benefits large-scale dataset generation or LLMs training. To enable apples-to-apples comparisons, we encourage future reports to specify (i) dataset-generation setup, (ii) training wall-clock and memory footprint, and (iii) per-sample inference cost.

\section{AI-assisted high-DoF metamaterial design}
\subsection{Image-based methods}
Most ML-based approaches for AI-assisted forward design of high-DoF metamaterials function as black-box models, where the metamaterial structure (geometries and material properties) serves as input, and the optical response (for instance, transmission and reflection coefficients) is predicted as output. In this context, ‘high-DoF’ refers to a hyper-dimensional design space characterized by independently tunable parameters (e.g., 10+), making traditional iterative method unfeasible. To encode the metamaterial unit cell into a format suitable for ML models, two primary digitalization methods have been adopted\citep{lee_datadriven_2024}. The first method involves pixelating the planar metamaterial and treating the resulting representation as an image, often in a binary format. Computer vision techniques, particularly CNNs, are then applied to optimize the pixel distribution. This approach allows for high-DoF designs, as each pixel can be independently adjusted to form intricate structures. Moreover, the pixel-based representation aligns with human intuition, as the unit cell's geometry is directly visualized, facilitating the interpretation of optimization outcomes. Research efforts have explored and refined this approach to enhance metamaterial design efficiency and performance\citep{fu_unleashing_2024,hegde_deep_2020,liu_nested_2022,majorel_deep_2022,singh_deep-learning_2024,teng_efficient_2023,han_inverse_2021,li_multifunctional_2024,zhu_multiplexing_2020,zhu_phase--pattern_2021,ma_probabilistic_2019,zhu_rapid_2024,liu_efficient_2024,li_predicting_2020,qu_convolutional_2023}.

One early demonstration of image-driven high-DoF metamaterial modelling was provided by An \textit{et al.} (2020),  who developed a CNN to predict wideband amplitude and phase responses of quasi-freeform dielectric metasurfaces\citep{an_deep_2020} (Fig.1 (a)). They leveraged the CNN model’s ability to handle structures across varying lattice constants, material indices, and thicknesses. They achieved an average prediction standard deviation of 0.005 (amplitude) and 0.78 degrees (phase) at each single frequency point after training on more than 100,000 simulation data sets. This study underscored the viability of image-based approaches to accelerate metamaterial designs, opening the door to fast performance evaluation for high-DoF structures. However, generating such an enormous training data set can be extremely time-consuming, raising concerns about the practical efficiency. One might question whether conventional simulation-optimization methods could achieve comparable performance with fewer simulations, calling into question the trade-off between model accuracy and data set size. Notably, while training required ~48 hours on two NVIDIA 1080 Ti GPUs, curating the >100k-sample corpus took ~8 days across six servers (~48 server-days), making data generation, not training, the dominant cost. Surrogate modelling only pays off when the model is reused extensively across many design queries; for one-off or small-batch tasks, the up-front data cost can outweigh the inference-time speedup.

This work was soon extended to exploit other network types. In 2021, An \textit{et al.} introduced a Generative Adversarial Network (GAN) for the inverse design of quasi-freeform structures\citep{an_multifunctional_2021} (Fig.1 (b)), using Wasserstein GAN (WGAN) that learns to generate free-form dielectric meta-atom patterns conditioned on desired amplitude and phase responses. They produced 100 qualified designs in 32 seconds under a tight threshold of $\pm$0.1 amplitude error and $\pm$10° phase error in dual-target cases. For the field of AI-assisted metamaterials design, this work marked a turning point by proving that GAN-based models can directly synthesize meta-atom layouts targeting multifunctional properties in one step. In 2022, Yu \textit{et al.} further improved An’s work by combining a Variational AutoEncoder (VAE) and GA for inverse design \citep{yu_inverse_2022} (Fig.1 (c)). Compared to GANs, the VAE compresses the large design space into a latent manifold of feasible structures, which acted as a search domain for a GA. By looping the GA over the VAE’s learned manifold of solutions, the algorithm could escape local optima and eventually converge to a meta-atom configuration that met the target spectral requirements. The training process lasts about 6h on two NVIDIA GeForce GTX 3080. In 2023, diffusion models entered the field: Zhang \textit{et al.} proposed a diffusion probabilistic model for generating high-DOF meta-atom images conditioned on desired wideband S-parameter spectra\citep{zhang_diffusion_2023} (Fig.1 (d)). Starting from random noise, the diffusion model iteratively refines the pixelated meta-atom design such that its simulated electromagnetic response converges to the specified target spectrum. This diffusion-driven strategy inherently avoids the training instabilities of GANs by eschewing adversarial objectives altogether. However, diffusion models require running additional sequential denoising steps to generate each design, which can significantly increase the computational time for inference. As a result, on-demand design generation is generally slower (0.43s per design as reported) compared to direct one-shot mapping techniques (such as those based on VAE or tandem networks, about 1ms per design on common commercial GPUs), as each solution must be iteratively computed, suggesting a trade-off between the speed for stability improvement and the accuracy in design outcomes. In 2025, Yang \textit{et al.} introduced a probabilistic generative model in a tandem architecture, i.e. the Tandem Generative Network (TGN), for design of meta-atoms\citep{yang_enhancing_2025} (Fig.1 (e)). The TGN architecture couples a forward neural network (to capture physics of a meta-atom’s response) with a generative network that samples new structure images from a learned probability space. This tandem setup addresses two key issues: the difficulty of handling one-to-many mappings in inverse problems (e.g., multiple structures yielding similar spectra) and the slow generation speed typical of vanilla diffusion models. Claiming up to 38\% lower mean absolute error (MAE)  and nearly 3000× faster generation (generated 10,000 atoms in 3.73s) than the diffusion model\citep{zhang_diffusion_2023}, TGN represents a further step in improving both speed and precision in high-DoF metasurface design.

Beyond quasi-freeform-related dielectric metasurfaces, several studies have applied image-processing networks to optimize other large, pixelated metasurfaces. For instance, Gahlmann and Tassin (2022) first trained a CNN to emulate the forward mapping from a 100×100 binary meta-atom image to its full spectra (S-parameters), then embedded this CNN into a conditional GAN (CGAN), which will propose new meta-atom images given target spectra\citep{gahlmann_deep_2022} (Fig.1 (f)). Similarly, Li \textit{et al.} (2022) proposed a CNN to predict the circular dichroism (CD) response of chiral metasurfaces with nanohole arrays (represented as 80×80 binary images)\citep{li_deep_2022}. In another approach, Tanriover \textit{et al.} (2022) introduced an AutoEncoder (AE) model for designing free-form 100×100 binary meta-atom images\citep{tanriover_deep_2022}. Collectively, these studies underscore the potential of DL to efficiently handle both forward and inverse design challenges while respecting fabrication constraints.

The rapid progress in image-based methods for metamaterial design points toward a future where scalability, interpretability, fabrication integration, and data efficiency become the focal points of research. On the scalability front, next-generation models will need to handle volumetric metamaterials as input “images”, which will enable the co-design of large-scale devices with many interacting elements without sacrificing resolution. Equally important is enhancing the interpretability of these models. As AI-designed metamaterials begin being put into practical use, designers will demand insights into how network features help to explain physical phenomena. Another critical direction is the seamless integration of manufacturing constraints and feedback into the design loop. Future AI models may incorporate differentiable fabrication-process simulators that ensure generated designs are not only nominally optimal but also robust against fabrication imperfections and material tolerances, which could involve training networks on experimental data. Additionally, emphasis on data efficiency will grow: instead of relying on tens of thousands of simulated examples, researchers are exploring physics-informed neural networks (PINNs), transfer learning, and active learning to make the most of limited data. We will cover the recent advances on these topics in our following discussions.

\begin{figure}
	\centering
	\includegraphics[width=\linewidth]{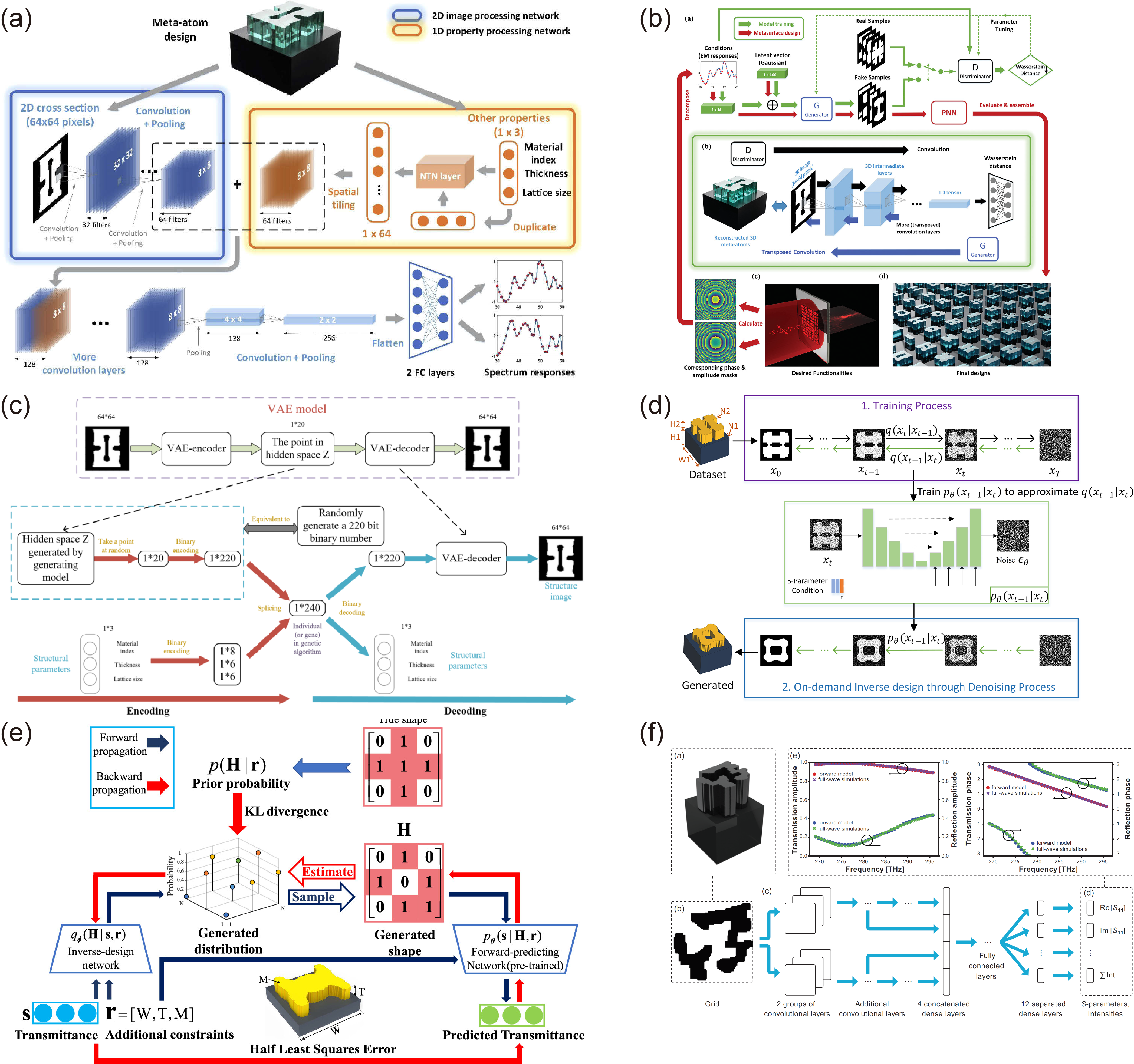} % pdf/png/eps
	\caption{Image-based method for AI-assisted design of metasurfaces. (a) CNN network for design of high-DoF quasi-freeform dielectric metasurfaces. Reprinted with permission from \citep{pillai_leveraging_2021}. Reprinted with permission from \citep{an_deep_2020}. Copyright 2020, Optical Society of America. (b) GAN model for metasurface inverse design47. Reprinted with permission from \citep{campbell_explosion_2021}. Copyright 2021, Wiley-VCH. (c) VAE and GA for metasurface inverse design. Reprinted with permission from \citep{park_free-form_2022}. Copyright 2022, Optical Society of America. (d) The first diffusion probabilistic model for inverse design of meta-atoms. Reprinted with permission from  \citep{zhang_diffusion_2023}. The article is licensed under a Creative Commons Attribution 4.0 International License. https://creativecommons.org/licenses/by/4.0/ (e) The first probabilistic generative model in a tandem architecture (TGN) for the design of meta-atoms. Reprinted with permission from \citep{yang_enhancing_2025}. Copyright 2025 American Chemical Society. (f) 100$\times$100 binary images were used for freeform metasurfaces forward and inverse design. Reprinted with permission from \citep{gahlmann_deep_2022}. Copyright 2022 American Physical Society. https://doi.org/10.1103/PhysRevB.106.085408
	}
\end{figure}

\subsection{Parameter-based methods}
While image-based methods have advanced the design of high-DoF planar metamaterials, their reliance on pixel discretization limits their implementation in true 3D meta-atoms with structural variations in the light propagation direction. To work around these constraints, parameter-based methods have been introduced as an alternative framework. In this approach, each meta-atom is represented by a vector that encodes its geometry and material properties, enabling a description of 3D subwavelength architectures. By shifting from a discrete pixel representation to a parameter space, these methods facilitate designs that capture intricate spatial details absent in image-based models. Sequence-to-sequence models, including FC layers and RNNs like Gated Recurrent Units (GRUs)\citep{cho_properties_2014} and LSTMs\citep{hochreiter_long_1997}, are often used to process these parameter vectors. The more recently emerged transformer architectures with self-attention mechanisms offered new possibilities for efficiently mapping complex parameter spaces to optical responses\citep{chen_broadband_2023}.

In their pioneering work published in 2017, Malkiel \textit{et al.} introduced a DL algorithm for plasmonic nanostructure design using an 8\mbox{-}parameter model to generate H\mbox{-}shaped planar structures within a design space comprising approximately $2.33\times 10^{8}$ possible configurations, addressing the long-standing challenge of time-consuming numerical simulation\citep{malkiel_deep_2017,malkiel_plasmonic_2018} (Fig.2 (a)). By training a bidirectional Deep Neural Network (DNN) composed of multiple FC layers on over 15,000 simulation instances, the authors achieved a transmission spectra prediction mean squared error (MSE) of 0.16. This study was significant as it was one of the first to solve a nontrivial metasurface design problem with DL, offering orders-of-magnitude speedups over iterative solvers. In 2018, Peurifoy \textit{et al.} showed that a DNN can approximate the forward light-scattering behaviours of multilayered nanoparticle structures with high accuracy using a relatively small training set \citep{peurifoy_nanophotonic_2018}(Fig.2 (b)). Once trained, their network reached a mean relative error (MRE) of 1.5\%. In a later study (2019), Nadell \textit{et al.} applied DL to model and design all-dielectric metasurfaces, which involve multiple resonant modes and near-field coupling between elements\citep{nadell_deep_2019} (Fig.2 (c)). Their method, which incorporated not only the raw parameters but also their ratios as inputs, achieved a transmittance prediction MSE of $1.16\times 10^{-3}$ after training on 18,000 simulation samples (about 0.1ms per prediction on Tesla Quadro M6000). This study validated that DNNs can handle relatively large, complex unit cells. In 2021, Xu \textit{et al.} combined NNs with transfer learning and GA to design phase-modulating metasurfaces\citep{xu_efficient_2021} (Fig.2 (d)). In their approach, a forward spectrum-prediction network was first trained on a base task (a rectangular meta-atom) and then fine-tuned (transferred) to a new task (elliptical meta-atom) using far fewer samples. The enhanced accuracy from transfer learning allowed the network to serve as a high-fidelity surrogate model for a genetic algorithm. The significance of this work lies in its hybrid strategy: by reducing the NN’s data requirements and integrating it with GA, it demonstrated a practical route to designing large-area functional metasurfaces more quickly. In 2022, Liao \textit{et al.} further extended AI-based design to 3D chiral plasmonic metasurfaces\citep{liao_deep_2022} (Fig.2 (e)). In practice, separate models were trained for a given handedness of a chiral structure; then the knowledge learned was partially transferred to a new model for the opposite handedness, greatly accelerating convergence while requiring little additional data. Looking to the current state of the art, researchers are pushing parameter-based methods to handle higher-dimensional design spaces and more complex unit cell architectures. In 2025, Zhang \textit{et al.} introduced a fixed-attention LSTM-based approach for both forward and inverse design of true 3D plasmonic metamaterials, defined by 12 parameters (representing two gold nanorods embedded in a dielectric substrate), thereby exploring a design space of approximately $3.09\times 10^{19}$ possible configurations (Fig.2 (f))\citep{zhang_fixed-attention_2025}. They treated the ordered list of design parameters as a temporal sequence and learned to “focus” to the most influential parameters during training. This attention-enhanced LSTM achieved ~48\% lower MSE on the metasurface’s transmission compared to a standard LSTM without attention, with about 3ms per prediction on one NVIDIA GeForce RTX 2080 Ti. This work addresses the “curse of dimensionality” in metamaterial design by intelligently structuring the network to handle many design variables (and their interdependencies), it opens the door for AI-assisted optimization of high-DoF metamaterials that were previously intractable.

In contrast to the studies mentioned above in this section, several recent studies have employed models based on a limited number of design parameters. For instance, Ma \textit{et al.} (2018) proposed a DL model for the design of stacked, twisted gold split ring resonators (SRRs) with dielectric spacers\citep{ma_deep-learning-enabled_2018} (Fig.2 (g)). After training on 25,000 samples, they achieved an MSE of $1.6\times 10^{-4}$ for reflection amplitude predictions. Liu \textit{et al.} (2018) directly confronted the one-to-many mapping issue inherent in photonic inverse design, where multiple distinct structures can exhibit the same spectrum\citep{liu_training_2018}(Fig.2 (h)). They introduced a tandem neural network training approach in which an inverse-design network is cascaded with a pretrained forward network (kept fixed) during training. Instead of learning an arbitrary mapping from spectrum to a particular geometry, the inverse model learns to produce any geometry that yields the desired spectrum by minimizing the error between the forward-predicted spectrum of its output design and the target spectrum. By effectively bypassing the need for one-to-one training pairs, this strategy enabled stable training of inverse models on datasets containing non-unique (i.e., degenerate) solutions, paving the way for reliable DL-based design of more complex photonic structures without being hindered by mode degeneracies or ambiguous mappings. Mall \textit{et al.} (2020) introduced a bidirectional AE (biAE) for plasmonic metasurfaces (with 4 structural parameters) that generated $10^{5}$ possible design configurations\citep{mall_fast_2020} (Fig.2 (i)), reaching an MAE of 1.43\% on validation cases after training on 1,200 full-wave simulation examples. Hou \textit{et al.} (2020) applied a tandem network for metamaterial absorbers defined by 6 parameters, achieving a test-set MSE of $2.95\times 10^{-4}$ after training on 20,000 samples (predicted in milliseconds per design on one NVIDIA GTX1060)\citep{hou_customized_2020} (Fig.2 (j)). Later studies by Han \textit{et al.} (2023) and Luo \textit{et al.} (2024) employed similar architectures (using 4 parameters) for designing chiral metastructures, with error metrics on the order of $10^{-4}$ (Fig.2 (k, l)) \citep{han_neural-network-enabled_2023,luo_flexible_2024}.

In Table~\ref{tab:table1}, we summarize the recent research on the degrees of freedom and the corresponding network architectures. For the “Dimensions” column, the categories are defined as follows: (i) 1D: meta-atom’s parameters vary along a single axis; (ii) 2D: meta-atoms can be represented as binary images. 2.5D: Meta-atoms can be described using a binary image combined with one additional parameter representing thickness information; (iii) 3D: true 3D meta-atoms with structural variations in the vertical direction. Although studies on 1D structures can efficiently explore a vast design space, many such systems can be rapidly solved using conventional optimization methods, suggesting the unnecessity of using AI-assisted approaches. Parameter-based AI design methods for metamaterials will evolve in several directions. First, as unit cells become more complex, future models must handle potentially dozens of design variables without compromising accuracy or requiring extensive training data. This evolution may involve network architectures such as attention mechanisms, physics-informed neural networks, and modular networks, along with dimensionality reduction techniques that isolate key design features. Transformer-based models and attention mechanisms offer a promising approach in this context. Also, multimodal learning that integrates electromagnetic simulations, experimental spectra, and fabrication constraints within a single model may further improve design efficiency. These advancements are expected to support the development of metamaterial systems that address practical design challenges and span high-dimensional design spaces.

\begin{figure}
	\centering
	\includegraphics[width=\linewidth]{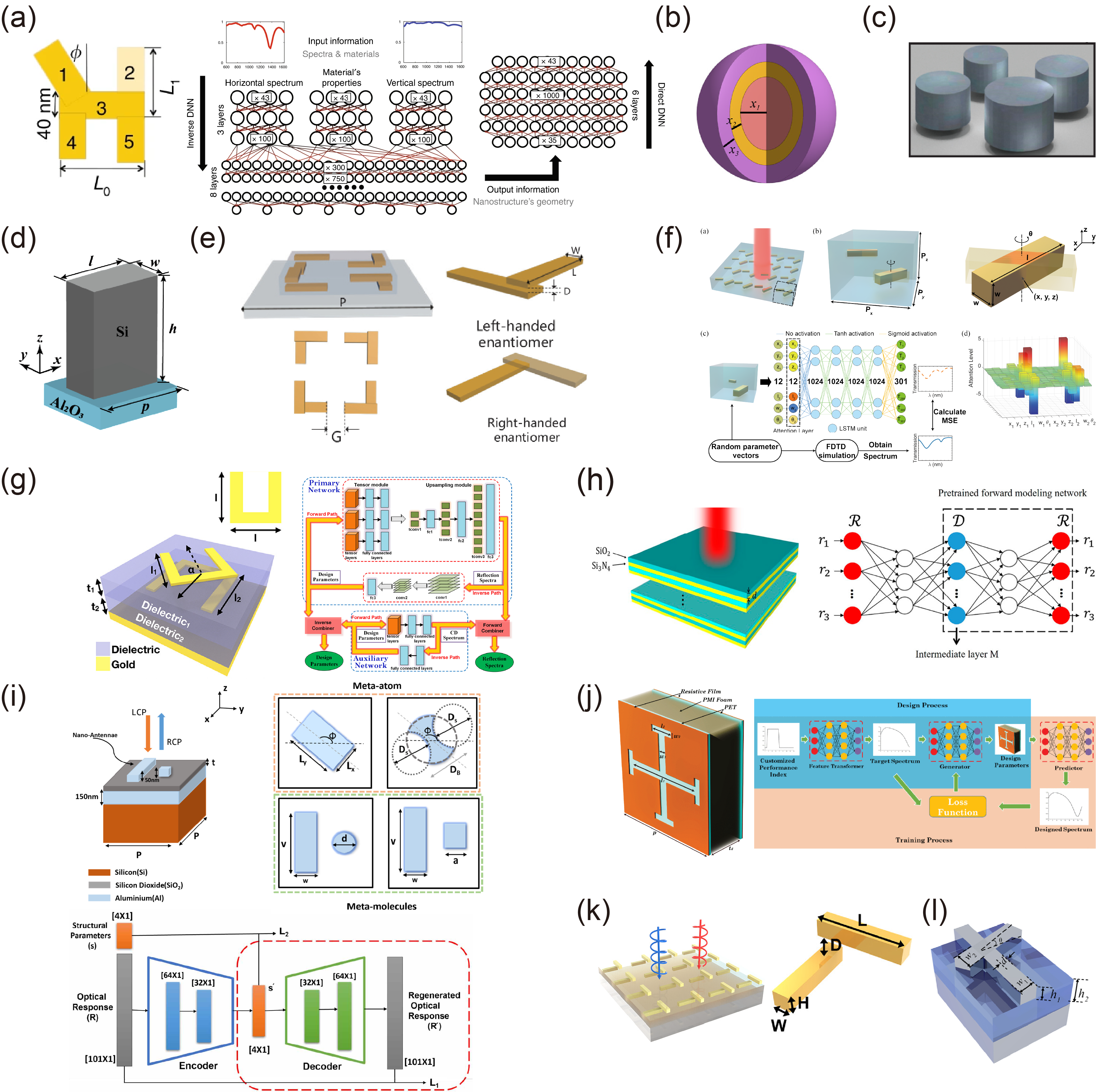} % pdf/png/eps
	\caption{Parameter-based method for AI-assisted design of metasurface. (a)  An H‑shaped planar structure design using DNN. Reproduced with permission from \citep{malkiel_plasmonic_2018}. TThe article is licensed under a Creative Commons Attribution 4.0 International License. https://creativecommons.org/licenses/by/4.0/. (b) Multilayered nanoparticle structures. Reproduced from \citep{peurifoy_nanophotonic_2018}. \copyright 2018 The Authors, with permission under the Creative Commons Attribution–NonCommercial 4.0 International License (CC BY-NC 4.0). http://creativecommons.org/licenses/by-nc/4.0/ (c) An all-dielectric metasurface with multiple resonant modes and near-field coupling between elements. Reprinted with permission from  \citep{nadell_deep_2019}. Copyright 2019, Optical Society of America. (d) A rectangular-shaped phase-modulating meta-structure. Reprinted with permission from \citep{xu_efficient_2021}. Copyright 2021, Optical Society of America. (e) A 3D Born-Kuhn type chiral metasurface. Reprinted with permission from \citep{liao_deep_2022}. Copyright 2022, Optical Society of America. (f) AI-assisted true 3D plasmonic high-DoF metamaterials design. Reproduced with permission from \citep{zhang_fixed-attention_2025}. Copyright 2025 Optical Society of America. (g) A stacked, twisted gold split ring resonator with dielectric spacers. Reproduced with permission from \citep{ma_deep-learning-enabled_2018}. Tandem networks were used to design thin-film metasurfaces. Reproduced with permission from \citep{liu_training_2018}. Copyright 2018 American Chemical Society. (i) Metal-dielectric-metal periodic gap-plasmon based half-wave plate metasurface design based on biAE. Reproduced with permission from \citep{mall_fast_2020}. Copyright 2020 Institute of Physics. (j) Metamaterial absorber design based on tandem networks. Reproduced with permission from  \citep{hou_customized_2020}. The article is licensed under a Creative Commons License. https://creativecommons.org/licenses/by/4.0/ (k) A chiral plasmonic Born–Kuhn metamaterial design based on multi-task learning. Reproduced with permission from  \citep{han_neural-network-enabled_2023}. Copyright 2023 American Chemical Society. (l) A dagger-shaped Ag array and an Ag mirror separated by a dielectric spacer. Reprinted with permission from  \citep{luo_flexible_2024}. Copyright 2024, Optical Society of America.}

\end{figure}

\begin{table*}[!htbp]
	\centering
	\small
	\setlength{\tabcolsep}{4.5pt}
	\renewcommand{\arraystretch}{1.12}
	\caption{Summary of recent parameter-based metasurface design methods. (Part 1 of 2)}
	\label{tab:table1}
	\begin{adjustbox}{max width=\textwidth, center}
		\begin{tabularx}{\textwidth}{Z Y Y Z Y Z Z}
			\toprule
			\makecell{\textbf{Works}} & \makecell{\textbf{Dim.}} & \makecell{\textbf{\#}\\\textbf{Params}} &
			\makecell{\textbf{Design Space}} & \makecell{\textbf{Train}\\\textbf{Set}} &
			\makecell{\textbf{Network Type}} & \makecell{\textbf{Error Metrics}} \\
			\midrule
			Malkiel \textit{et al.} (2017) \citep{malkiel_deep_2017} & 2D & 8 & $2.33\times 10^{8}$ * & 15{,}000 & FC layers & MSE: 0.16 \\
			Peurifoy \textit{et al.} (2018) \citep{peurifoy_nanophotonic_2018} & 1D & 8 & $7.98\times 10^{12}$ * & 50{,}000 & FC layers & MRE: 1.5\% \\
			Ma \textit{et al.} (2018) \citep{ma_deep-learning-enabled_2018} & 3D & 5 & Not mentioned & 25{,}000 & CNN & MSE: $1.6\times 10^{-4}$ \\
			Liu \textit{et al.} (2018) \citep{liu_training_2018} & 1D & 16 & $6.57\times 10^{34}$ * & Not mentioned & FC layers & Error: 0.19 \\
			Nadell \textit{et al.} (2019) \citep{nadell_deep_2019} & 2.5D & 8 & $8.16\times 10^{8}$ & 18{,}000 & FC layers & MSE: $1.16\times 10^{-3}$ \\
			An \textit{et al.} (2019) \citep{an_deep_2019} & 2.5D & 4 & $1.91\times 10^{11}$ * & 35{,}000 & FC layers & MSE: $3.5\times 10^{-4}$ \\
			Gao \textit{et al.} (2019) \citep{gao_bidirectional_2019} & 2.5D & 4 & $8.70\times 10^{8}$ & 3{,}900 & FC layers & MSE: $1.03\times 10^{-5}$ \\
			Lin \textit{et al.} (2019) \citep{lin_achieving_2019} & 2.5D & 4 & $4.12\times 10^{9}$ * & 25{,}900 & FC layers & MSE: $1.04\times 10^{-3}$ \\
			Li \textit{et al.} (2019) \citep{li_deep_2019} & 2.5D & 3 & $2.48\times 10^{7}$ * & 2{,}254 & FC layers & MSE: $3.86\times 10^{-5}$ \\
			Sajedian \textit{et al.} (2019) \citep{sajedian_double-deep_2019} & 2.5D & 8 & $5.72\times 10^{9}$ & Not mentioned & FC layers & Not applied \\
			Hou \textit{et al.} (2020) \citep{hou_customized_2020} & 2.5D & 6 & Not mentioned & 20{,}000 & FC layers & MSE: $2.95\times 10^{-4}$ \\
			Unni \textit{et al.} (2020) \citep{unni_deep_2020} & 1D & 11 & $1.79\times 10^{26}$ * & 100{,}800 & FC layers & Negative log-likelihood: -4.5 \\
			Mall \textit{et al.} (2020) \citep{mall_fast_2020} & 2D & 4 & $1.00\times 10^{5}$ & 1{,}200 & biAE & MAE: 1.43\% \\
			Tanriover \textit{et al.} (2020) \citep{tanriover_physics-based_2020} & 2.5D & 3 & $1.41\times 10^{8}$ * & 3{,}157 & FC layers & MSE: $2.1\times 10^{-3}$ \\
			Qiu \textit{et al.} (2021) \citep{qiu_simultaneous_2021} & 1D & 8 & $2.03\times 10^{7}$ * & 80{,}000 & FC layers & MSE: $5\times 10^{-2}$ \\
			Xu \textit{et al.} (2020) \citep{xu_enhanced_2020} & 2D & 3 & Not mentioned & 25{,}000 & FC layers & MSE: $5\times 10^{-3}$ \\
			Unni \textit{et al.} (2021) \citep{unni_mixture-density-based_2021} & 1D & 20 & $4.30\times 10^{39}$ * & 579{,}600 & FC layers & RMSE: $2\times 10^{-2}$ \\
			Xu \textit{et al.} (2021) \citep{xu_efficient_2021} & 2.5D & 4 & $9.29\times 10^{8}$ * & 27{,}000 & FC layers & MSE: $7.7\times 10^{-4}$ \\
			Lininger \textit{et al.} (2021) \citep{lininger_general_2021} & 1D & 5 & $1.00\times 10^{12}$ & 200{,}000 & CNN & RMSE: $2\times 10^{-2}$ \\
			Zandehshahvar \textit{et al.} (2021) \citep{zandehshahvar_inverse_2021} & 1D & 8 & $7.98\times 10^{12}$ * & 40{,}000 & AE & MSE: $2.2\times 10^{-6}$ \\
			Huang \textit{et al.} (2021) \citep{huang_inverse_2021} & 2D & 5 & $3.03\times 10^{14}$ * & 12{,}040 & FC layers & MSE: $1.4\times 10^{-2}$ \\
			Sun \textit{et al.} (2021) \citep{sun_machine_2022} & 2D & 4 & $5.94\times 10^{7}$ * & $2.16\times 10^{7}$ & K-nearest neighbor (KNN) & MSE: $3.46\times 10^{-6}$ \\
			Tanriover \textit{et al.} (2021) \citep{tanriover_neural_2021} & 2.5D & 4 & $1.61\times 10^{11}$ * & 6{,}318 & Complex valued FC layers & MSE: $1.2\times 10^{-4}$ \\
			Deng \textit{et al.} (2021) \citep{deng_neural-adjoint_2021} & 2.5D & 14 & $1.04\times 10^{12}$ & 24{,}000 & CNN & MSE: $1.2\times 10^{-3}$ \\
			\bottomrule
		\end{tabularx}
	\end{adjustbox}
\end{table*}

\begin{table*}[!htbp]
	\centering
	\small
	\setlength{\tabcolsep}{4.5pt}
	\renewcommand{\arraystretch}{1.12}
	\ContinuedFloat
	\caption*{\textbf{Table \ref{tab:table1} (continued)}}
	\begin{adjustbox}{max width=\textwidth, center}
		\begin{tabularx}{\textwidth}{Z Y Y Z Y Z Z}
			\toprule
			\makecell{\textbf{Works}} & \makecell{\textbf{Dim.}} & \makecell{\textbf{\#}\\\textbf{Params}} &
			\makecell{\textbf{Design Space}} & \makecell{\textbf{Train}\\\textbf{Set}} &
			\makecell{\textbf{Network Type}} & \makecell{\textbf{Error Metrics}} \\
			\midrule
			Xu \textit{et al.} (2021) \citep{xu_improved_2021} & 2D & 4 & $8.55\times 10^{7}$ * & 71{,}808 & FC layers & Accuracy: 96.49\% \\
			Noureen \textit{et al.} (2022) \citep{noureen_unique_2022} & 2.5D & 7 & $7.39\times 10^{12}$ * & Not mentioned & FC layers & MSE: $1.8\times 10^{-3}$ \\
			Liao \textit{et al.} (2022) \citep{liao_deep_2022} & 2.5D & 5 & $4.25\times 10^{7}$ * & 23{,}000 & FC layers & MSE: $1.6\times 10^{-4}$ \\
			Gao \textit{et al.} (2022) \citep{gao_deep-learning-assisted_2022} & 2D & 4 & $2.08\times 10^{12}$ * & 10{,}350 & Modified FC layers & MSE: $1.47\times 10^{-4}$ \\
			Shen \textit{et al.} (2022) \citep{shen_inverse_2022} & 2.5D & 4 & $3.06\times 10^{6}$ & 6{,}000 & FC layers & MSE: $1.02\times 10^{-3}$ \\
			Deng \textit{et al.} (2022) \citep{deng_long_2022} & 2.5D & 9 & $9.00\times 10^{12}$ & 18{,}000 & biRNN & MSE: $1.34\times 10^{-3}$ \\
			Lin \textit{et al.} (2022) \citep{lin_high-performance_2022} & 2.5D & 3 & $2.94\times 10^{11}$ & 2{,}322 & FC layers & MSE: $2.19\times 10^{-4}$ \\
			Li \textit{et al.} (2022) \citep{li_neural_2022} & 2.5D & 3 & $7.59\times 10^{8}$ * & 42{,}000 & FC layers & MSE: $1.06\times 10^{-3}$ \\
			Knightley \textit{et al.} (2023) \citep{knightley_neural_2023} & 2.5D & 4 & $4.05\times 10^{7}$ & 9{,}360 & FC layers & MSE: $4.70\times 10^{-3}$ \\
			Chen \textit{et al.} (2022) \citep{chen_prediction_2022} & 2.5D & 20 & Not mentioned & 23{,}000 & CNN & MSE: $2.3\times 10^{-3}$ \\
			Qiu \textit{et al.} (2023) \citep{qiu_chiral_2023} & 2.5D & 4 & $1.60\times 10^{7}$ & 12{,}000 & FC layers & MSE: $1.57\times 10^{-3}$ \\
			Liu \textit{et al.} (2023) \citep{liu_deep_2024} & 2.5D & 4 & $1.80\times 10^{8}$ & 50{,}000 & FC layers & MSE: $4.8\times 10^{-3}$ \\
			Jiang \textit{et al.} (2023) \citep{jiang_generic_2024} & 2.5D & 5 & $4.05\times 10^{7}$ & 201 & CNN & MSE: $5.19\times 10^{-4}$ \\
			Yu \textit{et al.} (2023) \citep{yu_hybrid_2023} & 2.5D & 2 & $3.40\times 10^{6}$ & 640 & FC layers & MSE: $2.43\times 10^{-4}$ \\
			Han \textit{et al.} (2023) \citep{han_neural-network-enabled_2023} & 2.5D & 4 & $1.60\times 10^{7}$ & 5{,}628 & MTL-FC layers & MSE: $1.12\times 10^{-4}$ \\
			Jahan \textit{et al.} (2024) \citep{jahan_deep_2024} & 2.5D & 11 & $1.10\times 10^{14}$ & Not mentioned & FC layers & MSE: $5.0\times 10^{-3}$ \\
			Luo \textit{et al.} (2024) \citep{luo_flexible_2024} & 2.5D & 4 & $2.40\times 10^{7}$ & 12{,}000 & CNN & MSE: $8.03\times 10^{-4}$ \\
			Chen \textit{et al.} (2024) \citep{chen_inverse_2024} & 2.5D & 3 & $4.05\times 10^{7}$ & 5{,}000 & MLP & MSE: $1.638\times 10^{-2}$ \\
			Wang \textit{et al.} (2024) \citep{wang_inverse_2024} & 2.5D & 3 & $4.05\times 10^{7}$ & 5{,}000 & MLP & MSE: $1.638\times 10^{-2}$ \\
			Zhu \textit{et al.} (2024) \citep{zhu_optimized_2024} & 2D & 9 & $3.47\times 10^{13}$ $^{*}$ & 4{,}736 & FC layers & MSE: $6\times 10^{-4}$ \\
			Fan \textit{et al.} (2024) \citep{fan_prediction_2024} & 2.5D & 5 & Not mentioned & 2{,}400 & FC layers & MSE: $7.4\times 10^{-3}$ \\
			Liu \textit{et al.} (2025) \citep{liu_cascaded_2025} & 2.5D & 6 & $8.90\times 10^{6}$ $^{*}$ & 12{,}988 & FC layers & MSE: $3\times 10^{-5}$ \\
			Yu \textit{et al.} (2025) \citep{yu_prediction_2025} & 2.5D & 5 & Not mentioned & 2{,}400 & FC layers & MSE: $1\times 10^{-3}$ \\
			Chen \textit{et al.} (2025) \citep{chen_thermal_2025} & 2.5D & 16 & $1.00\times 10^{43}$ & Not mentioned & FC layers + Bayesian linear regression & Not mentioned \\
			Zhang \textit{et al.} (2025) \citep{zhang_fixed-attention_2025} & 3D & 12 & $3.09\times 10^{19}$ & 6{,}393 & LSTM with Fixed-Attention & MSE: $2.17\times 10^{-3}$ \\
			\bottomrule
		\end{tabularx}
	\end{adjustbox}
\end{table*}

% Keep the tables with their section:
\FloatBarrier

\section{Transformers and attentions applied in metamaterial design}
Transformers are a modern DL architecture that relies on an attention mechanism to process information, rather than the convolution or recurrence used in traditional models (Fig.3 (a)) \citep{vaswani_attention_2017}. When computing an output, transformers ingest the entire input (e.g., a sentence or an image) and use self-attention to decide which parts of the input are most relevant to each other. In other words, the model learns to weigh the influence of different input elements on each other dynamically. This attention-driven approach allows transformers to capture long-range dependencies in data effectively – for example, a word at the beginning of a sentence can directly influence the interpretation of a word at the end, because the model can access and evaluate both simultaneously. The ability to look at all parts of the input at once is also essential to mitigate issues including the vanishing-gradient problem that RNNs face when dealing with long sequences. Transformers have since become the dominant model in natural language processing and are increasingly used in other domains (with variants such as the Vision Transformers for images) \citep{han_survey_2023}.

The application of transformers has recently expanded to the design of metamaterials. In 2023, Chen \textit{et al.} introduced the first encoder-only transformers for both forward and inverse design of broadband solar metamaterial absorbers (Fig.3 (b)) \citep{chen_broadband_2023}. The studied absorbers comprise 6 subwavelength layers with thicknesses ranging from 0 to 100 nm for Ge, Si, TiO$_2$, Al$_2$O$_3$, SiO$_2$ and 0–200 nm for MgF$_2$. To handle the high-dimensional spectral data, the input spectrum is segmented into multiple patches. Each patch is embedded using one-dimensional convolution before being fed into a transformer encoder. This segmentation and positional embedding help overcome overfitting and dimension mismatch issues, enabling the network to learn the underlying physical relationships effectively, and is expected to be implemented frequently in future works. After training, prediction is millisecond-scale on one NVIDIA GeForce GTX 3080 Ti. This work represents a significant breakthrough in AI-assisted design of metamaterials incorporating transformers, a new paradigm in navigating complex optical design spaces with unprecedented efficiency and accuracy. In the next study, Chen \textit{et al.} (2023) have extended their approach to design a dielectric metasurface where the incident angle is tuned to mediate the coupling between guided-mode resonances and quasi-BICs (Fig.3 (c)) \citep{chen_alldielectric_2024}. With unit cells characterized by 5 parameters and a training set containing 23,681 simulation data points, their encoder-only transformers achieved an MSE of $4.56\times10^{-3}$, which represents a 17.6\% improvement over a FC layers network, and a 34.4\% reduction in training parameters. Other encoder-only approaches have also emerged. For example, Niu \textit{et al.} (2023) have employed a VAE with an encoder-only transformer for the inverse design of metasurfaces \citep{niu_deep_2023}, while Yin \textit{et al.} (2025) have combined a transformer encoder for forward prediction (~17 ms per prediction) with a visual attention network for inverse design \citep{yin_deep_2025}. In another study, Ma \textit{et al.} (2025) have integrated FC layers with a transformer to achieve the inverse design of a metasurface absorber \citep{ma_trmd_2025}. These models leverage the transformer encoder’s ability to capture long-range dependencies within metamaterial data in parallel, making them well-suited for both forward and inverse tasks.

Some studies have implemented other transformer architectures \citep{gao_metaattention_2024}. For instance, Huang \textit{et al.} (2024) have applied an improved transformer combined with a CGAN for the inverse-design of graphene terahertz multi-resonant metasurfaces, represented by a 20-element chemical potential vector (Fig.3 (d)) \citep{huang_artificial_2024}. After trained on 19,000 data points, the network achieved a test accuracy of 96.14\% compared to 94.27\% for an FC-layer model. In a separate effort, Ma \textit{et al.} (2024) implemented a decoder-only transformer, which is also known as a Generative Pre-trained Transformer (GPT), for inverse design of multilayer thin film structures (Fig.3 (e)) \citep{ma_optogpt_2024}. Rather than fixing the material for each layer and optimizing only the thickness, they introduced a “structure token” that defines both the material and its thickness, such as “Al\_10”. This approach not only overcomes the limitations of fixed output sizes but also enables handling diverse design scenarios, including variable numbers of layers, distinct material combinations, and different incidence angles and polarization states.

Despite these advances, transformers generally require large data sets because of their limited inductive biases \citep{liu_efficient_2021}. Early studies noted that RNN-based sequence-to-sequence models could match or exceed the performance of transformers on small parallel data sets, underscoring the training challenges when data is limited \citep{lankford_transformers_2024}. More recent work has demonstrated that careful hyperparameter tuning, regularization, and strategies such as reducing network depth, employing smaller token vocabularies, or leveraging pre-training and transfer learning can help transformers perform in low-resource settings \citep{lankford_transformers_2024}. Therefore, we believe that relaxing the data requirements of transformers remains an important area for future research. Notwithstanding the remarkable achievements of Vision Transformers in computer vision tasks, their potential for image-based metamaterial design remains largely untapped, presenting a promising avenue for future research.

Beyond transformer architectures, incorporating self-attention into alternative network structures has also boosted the efficiency of metamaterial design. For example, Zeng \textit{et al.} (2023) have investigated the data shift in electromagnetic solvers for 1D grating couplers by integrating a ResNet with mixed training and multihead attention \citep{zeng_utilizing_2023}. They observed that when models are trained on data based on randomly generated nano-structures but then applied to predict optimized designs, a mismatch in data distributions (data shift) causes a significant drop in prediction accuracy. They also introduced a mixed training strategy, where a small fraction of the optimized (shifted) data is blended into the training set. The reported model achieved an MSE of $1.35\times10^{-4}$ for coupling efficiency prediction, marking an improvement by 167\% compared to a ResNet without attention. Similarly, Yuan \textit{et al.} (2024) have inversely-designed tunable mid-infrared metasurface-embedded Fabry-Perot filters via FC layers that couple with a CNN-self-attention module for forward modelling (Fig.3 (f)) \citep{yuan_multitask_2024}. Trained on 19,473 simulation samples, this approach yielded a test R$^2$ of 0.973 for predicting transmission as defined in their paper, approximately 10\% higher than a comparable network without self-attention. These studies demonstrate the potential of self-attention mechanisms in the design of metamaterials. In particular, the reported results confirm that integrating self-attention mechanisms can reduce prediction errors and enhance model accuracy. Further investigation into such integrations, which lead to systems typically requiring less training data than transformers, is expected to advance the overall capabilities of AI-assisted metamaterial design.

\begin{figure}
	\centering
	\includegraphics[width=\linewidth]{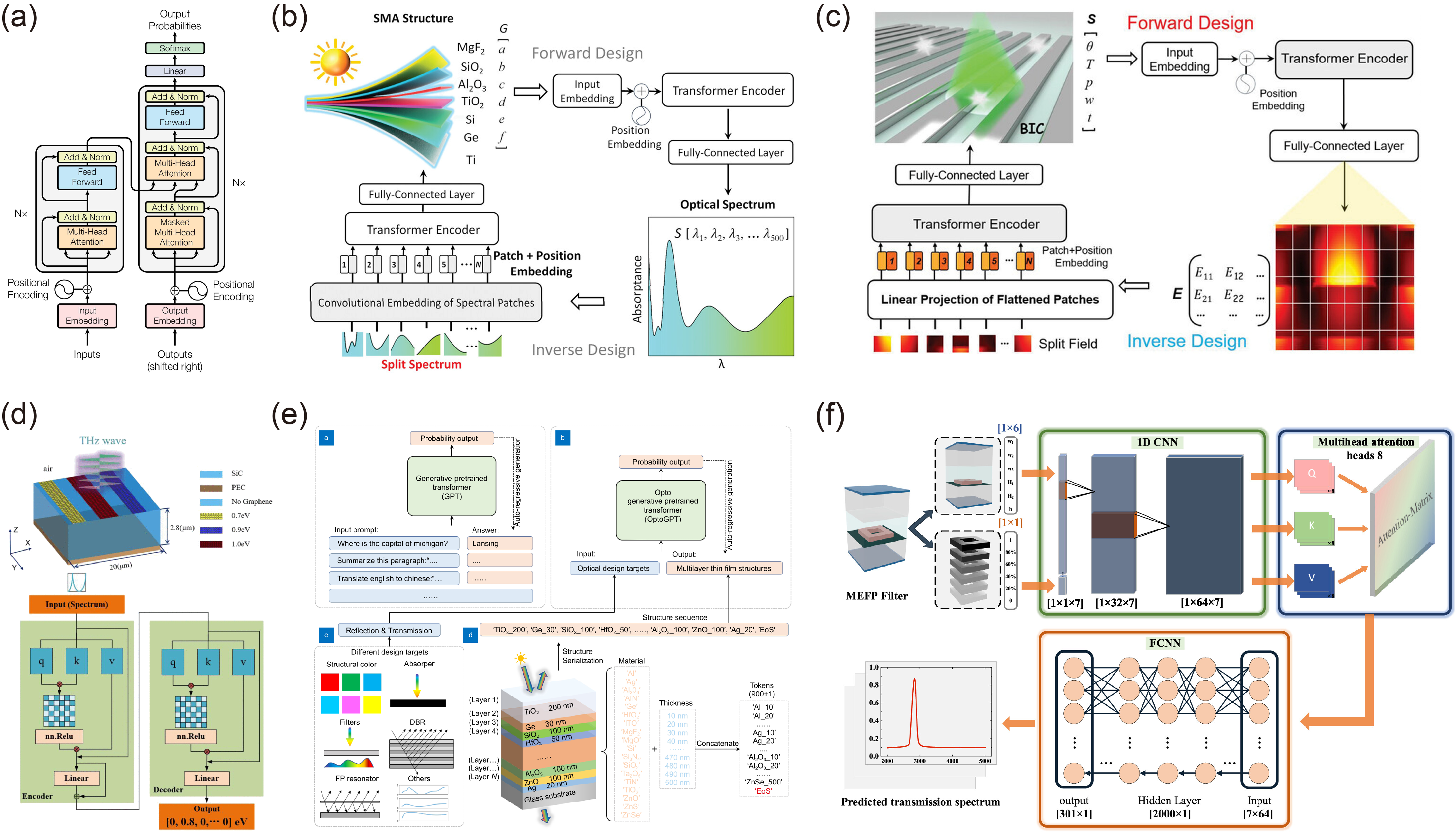} % pdf/png/eps
	\caption{Transformer and self-attention for AI-assisted design of metasurfaces. (a) Transformer architecture. Reproduced from \citep{vaswani_attention_2017} with permission from Google, which grants reproduction of tables and figures for scholarly works provided proper attribution is given. (b) Encoder-only transformers for the design of broadband solar metamaterial absorbers. Reproduced from \citep{chen_broadband_2023} Copyright 2023 The Authors, Advanced Photonics Research published by Wiley-VCH GmbH, under the terms of the Creative Commons Attribution License. https://creativecommons.org/licenses/by/4.0/ (c) Dielectric metasurface design based on encoder-only transformer models. Reproduced from \citep{chen_alldielectric_2024} Copyright 2023 The Authors, Advanced Optical Materials published by Wiley-VCH GmbH (d) Improved transformer combined with a CGAN for the inverse-design of graphene terahertz multi-resonant metasurfaces. Reproduced with permission from \citep{huang_artificial_2024}. Copyright 2023 IEEE.  (e) GPT for inverse design of multilayer thin film structures. Reprinted with permission from \citep{ma_optogpt_2024}. The article is licensed under a CC-BY 4.0 License. https://creativecommons.org/licenses/by/4.0/ (f) Mid-infrared metasurface-embedded Fabry-Perot filters design via FC layers that couple with a CNN-self-attention module. Reproduced with permission from \citep{yuan_multitask_2024}. Copyright 2024 American Chemical Society.}

\end{figure}

\section{Prediction of mutual coupling effects}
Mutual coupling refers to the complex electromagnetic interactions between closely spaced meta-atoms, which can significantly alter the idealized responses assumed during conventional design\citep{liu_coupling_2010}. Traditional design methods often apply periodic boundary conditions or approximations that neglect near-field interactions, which can lead to discrepancies between unit-cell based simulation results and actual device performance, particularly for phase-sensitive metasurfaces utilized for wavefront manipulation\citep{olk_accurate_2019}. Modelling these interactions using full-wave electromagnetic simulations demands extensive computational resources, thereby hindering efficient design optimization. AI-based methods address this challenge by learning the mapping between meta-atom geometry, local environment, and optical response\citep{fan_prediction_2024,zhelyeznyakov_deep_2021,ma_deep_2024,bao_gat-net_2025,li_deep-learning-based_2023}.

In 2021, An and colleagues developed a CNN to accurately predict the electromagnetic responses of individual meta-atoms when mutual coupling between nonidentical neighbours is present (Fig.4 (a))\citep{an_deep_2022}. The core idea involved translating the physical configurations of the target and adjacent meta-atoms into high-resolution, binarized images, where dielectric regions and voids were distinctly marked, then processing these images through the CNN to extract key spatial features. When integrated with a global optimization, the method increased a beam deflection efficiency from 41.3\% to 68.8\% and improved a meta-lens's focusing efficiency by over 20\%. In addition, An \textit{et al.} demonstrated that by accounting for coupling perturbations, devices such as beam deflectors and meta-lenses exhibit significantly enhanced efficiency, ensuring that a larger fraction of the incident energy is effectively manipulated. In 2022, Majorel \textit{et al.} developed a U-Net-based CNN to model optical responses of complex, aperiodic plasmonic metasurfaces that can extend to arbitrarily large sizes (Fig.4 (b))\citep{majorel_deep_2022}. Instead of repeatedly calculating detailed optical interactions for every configuration, the authors proposed to approximate a “dressed polarizability” for each nanostructure. This quantity encapsulates how local interactions (due to neighboring coupling, substrates, and other environmental factors) alter the response of an individual nanostructure. The CNN takes a 2D top-view image of the nanostructure arrangement along with wavelength information as input and is trained to output the complex-valued dressed polarizability tensor for each nanostructure. This method provided scalability to arbitrarily large and complex geometries for future references. In 2023, Ma \textit{et al.} proposed a DL model for rapid calculation and optimization of metasurfaces incorporating meta-atom interactions by training separate network blocks to associate reflection phase and amplitude with specific meta-atoms (Fig.4 (c)) \citep{ma_incorporating_2023}. Rather than treating the DNN as a black box, their approach interprets weight values in cascaded dense layers as representing physical mechanisms of electromagnetic scattering. In the same year, Ha \textit{et al.} developed a DL optimizer for large-aperture meta-lens design that segments the lens into overlapping 5×5 super meta-atoms to capture local lattice interactions (Fig.4 (d))\citep{ha_physics-data-driven_2023}. Their architecture integrates an AE to extract low-dimensional representations of geometrical features with an inverse-design network that refines meta-atom dimensions to mitigate coupling-induced phase errors, resulting in a fabricated meta-lens with a 1 mm radius and a relative focusing efficiency of 93.4\% (compared to the ideal focusing efficiency).

Moving forward, research on AI-assisted metasurface design is expected to further embrace models that inherently capture meta-atom responses over varying length scales while maintaining physical interpretability. Emerging network architectures such as transformers offer the ability to learn long-range dependencies across an entire metasurface, using self-attention or fixed-attention to account for both local and global coupling effects without a preset neighbor limit. Equally important is infusing more physics knowledge into the learning process to avoid purely black-box behaviour. This could involve embedding constraints like energy conservation, reciprocity, or known coupling formulas into model architectures or loss functions, as well as designing network outputs that correspond to physically meaningful parameters.

\begin{figure}
	\centering
	\includegraphics[width=\linewidth]{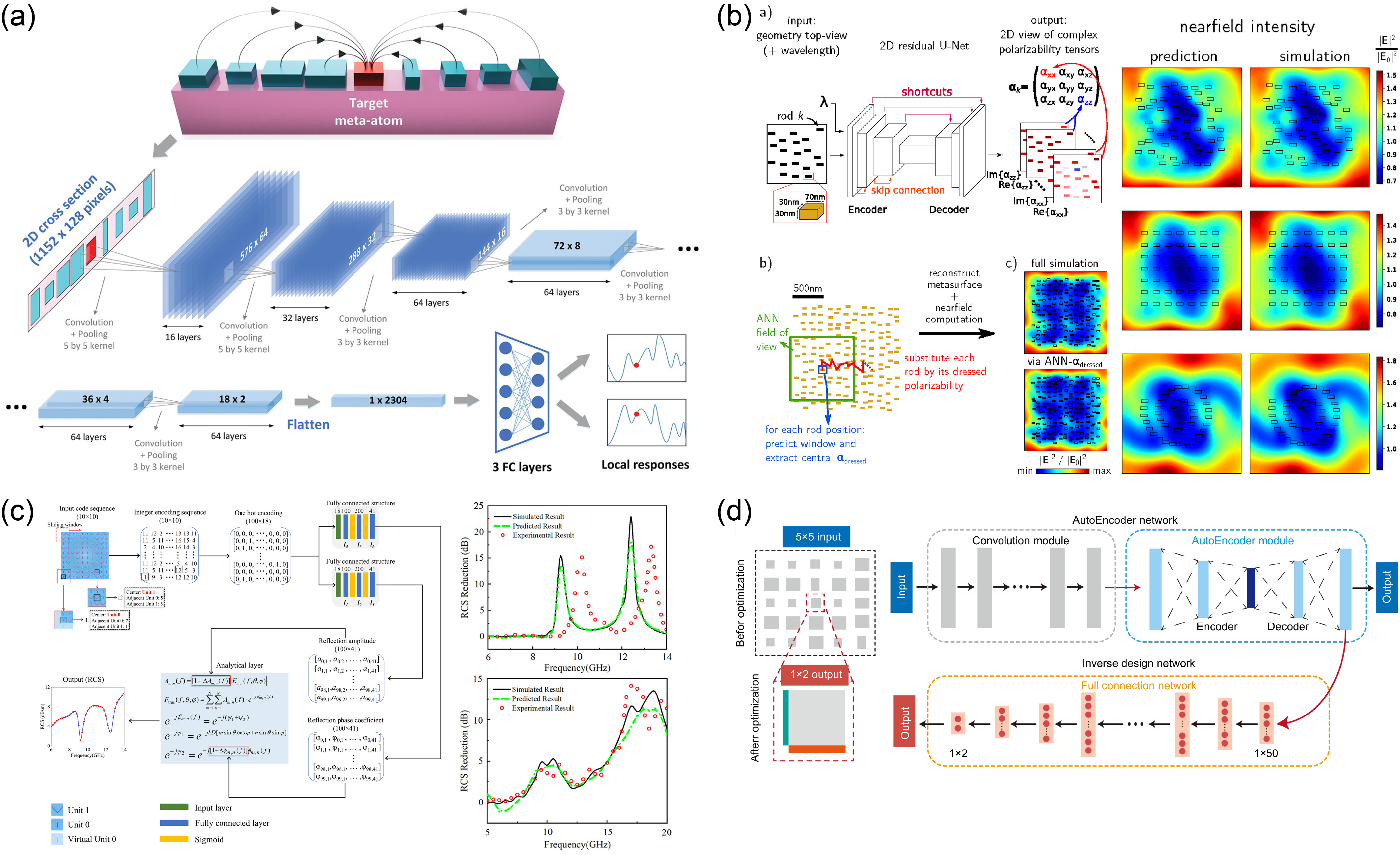} % pdf/png/eps
	\caption{AI-assisted design of metasurfaces with mutual coupling effects. (a) Prediction of electromagnetic responses of individual meta-atoms when mutual coupling between nonidentical neighbours is present via CNN. Reproduced from \citep{an_deep_2022} Copyright 2021 The Authors, Advanced Optical Materials published by Wiley-VCH GmbH (b) U-Net-based CNN for the modelling of complex, aperiodic plasmonic metasurfaces that can extend to arbitrarily large sizes. Reproduced with permission from \citep{majorel_deep_2022}. Copyright 2022 American Chemical Society. (c) Rapid calculation and optimization of metasurfaces incorporating meta-atom interactions. Reproduced from \citep{ma_incorporating_2023}. Copyright 2023 The Authors, Advanced Photonics Research published by Wiley-VCH GmbH, under the terms of the Creative Commons Attribution License. https://creativecommons.org/licenses/by/4.0/ (d) A DL optimizer for large-aperture meta-lens design via AE. Reprinted with permission from \citep{ha_physics-data-driven_2023}. The article is licensed under a CC-BY 4.0 License. https://creativecommons.org/licenses/by/4.0/}
	
\end{figure}

\section{Robust and fabrication-friendly metamaterial design}
DNNs are making otherwise intractable problems a reality. In inverse-design, one typically seeks to maximize a set of performance criteria given a range of input parameter values. While these ranges may correspond to available material values or geometrical dimensions, such constraints are not the same as tolerances. In fact, tolerances are usually not considered in situ during optimization, but rather a posteriori, if at all. This means that the optimizer has no idea about the sensitivity of the response surface (i.e., the hyper-dimensional objective space) to changes in input parameters. It is actually quite possible that the optimizer, simply seeking to maximize nominal performance, finds a solution that is highly-performant, but quite sensitive to input uncertainties. Designers therefore need to have previous experience when analysing optimized designs to anticipate their potential sensitivity. Otherwise, tolerance analysis may be performed using conventional Monte Carlo methods to estimate a design’s guaranteed minimum performance given a set of input tolerances/uncertainties. However, in the case of computationally expensive models such as those requiring full-wave analysis or those with many input dimensions, this kind of analysis itself can be prohibitive. To this end, engineers have used surrogate modelling approaches based on radial basis functions or the Kriging model to help with such analysis. To this end, Easum \textit{et al.} (2018) \citep{easum_efficient_2018} introduced an optimization algorithm that iteratively trains surrogate models which accurately capture response surface features in order to calculate a design’s “tolerance hypervolume.” Due to the multi-objective nature of the algorithm, it presents designers with a Pareto front of optimized designs that showcase the trade-offs between performance objectives and robustness. While this technique represented a breakthrough in RF antenna optimization, classical surrogate models aren’t equipped to handle the more sophisticated nonlinear relationships seen in nanophotonic meta-devices. Therefore, new techniques were needed in order to capture design robustness in freeform optical meta-devices.

Wen \textit{et al.} demonstrated in 2020 how a progressively growing GAN (PGGAN) can learn how to output highly efficient and robust metasurfaces using only a sparse training data set (Fig.5 (a)) \citep{wen_robust_2020}. The progressive growth aspect of the network enabled more robust learning of local topological features in the training set while a self-attention mechanism allowed the network to capture global features. The PGGAN was able to significantly accelerate the process for producing robust optimizes designs compared to the conventional topology (i.e., adjoint) optimization process. In 2021, Jenkins \textit{et al.} demonstrated a U-Net based architecture that achieved extremely accurate prediction of a metasurface’s performance under geometric variations (Fig.5 (b)) \citep{jenkins_establishing_2021}. The network was used in conjunction with a more conventional multi-objective optimization algorithm in order to quantify both nominal (i.e., no variations) and guaranteed minimum performances over all possible geometric deviations. Computing the guaranteed minimum performance is difficult as it requires an exhaustive evaluation of all possible variations; the minimum performance does not always occur at the extremes of the uncertainty range. Using this approach, the authors presented a result that trades off a few percent of nominal performance for over a 100\% increase in guaranteed minimum performance. Moreover, the hybrid DL approach reduced the inverse-design process from a potential many months’ time scale to that of just a few days, overall speedups 14.8 times for a single optimization ignoring startup time and 4.37 times including it.

In 2023, Tanriover \textit{et al.} combined an AE and FC NN to produce a DL model capable of generating manufacturable freeform dielectric meta-atoms (Fig.5 (c)) \citep{tanriover_deep_2022}. Their approach sought to improve upon model generalizability and fabrication feasibility compared to other solutions. The forward model exhibited generalizability in material dispersion, source polarization, and wavelength range of operation and was subsequently connected to a GA to perform design optimization. In the same year, Ueno \textit{et al.} presented a DL framework for producing fabrication-friendly metasurfaces. The authors demonstrated a dual-band optical collimator whose inverse-design was accelerated by the DNN (Fig.5 (d)) \citep{ueno_dual-band_2023}.The free-form meta-atoms were generated using a predictive neural network (PNN) which was trained to accurately predict the transmission phase and amplitude of candidate designs.

\begin{figure}
	\centering
	\includegraphics[width=\linewidth]{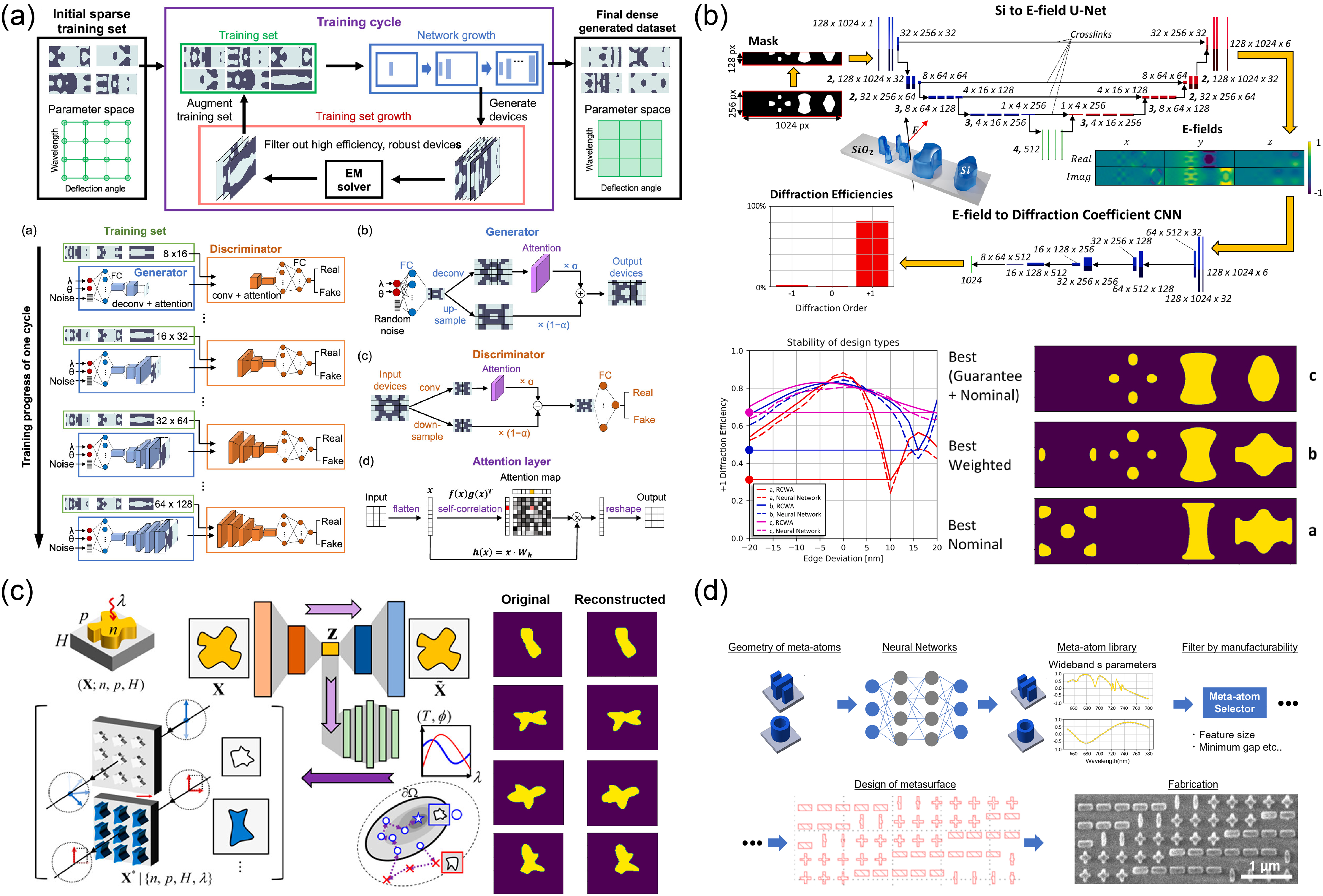} % pdf/png/eps
	\caption{AI-assisted design of robust and fabrication-friendly metasurfaces. (a) PGGAN with self-attention rapidly output freeform metasurface designs that surpass topology-optimized devices in efficiency and robustness. Reproduced with permission from \citep{wen_robust_2020}. Copyright 2020 American Chemical Society. (b) A U-net based DNN with evolutionary optimization to design metasurfaces whose efficiency persists across fine-grained fabrication-induced edge deviations. Reprinted with permission from \citep{jenkins_establishing_2021}. The article is licensed under a Creative Commons Attribution 4.0 International License. https://creativecommons.org/licenses/by/4.0/ (c) An end-to-end generative-modelling pipeline that learns manufacturable free-form dielectric metasurface shapes. Reproduced with permission from \citep{tanriover_deep_2022}. Copyright 2022 American Chemical Society. (d) A DL-generated, fabrication-constrained library of free-form meta-atoms for the design of metasurface collimators. Reprinted with permission from \citep{ueno_dual-band_2023}. The article is licensed under a Creative Commons Attribution 4.0 International License. https://creativecommons.org/licenses/by/4.0/.}
	
\end{figure}

\section{Recent Advancements and Perspectives}
Recent research highlights several emerging trends that are shaping the future of AI-assisted metamaterial design. These approaches aim to overcome current limitations (such as data scarcity, limited generalization, or design complexity) and open new possibilities for metamaterial engineering. 

One emerging approach is the hybridization of different neural network types within a single design framework. The motivation is that complex metamaterial design tasks often involve multiple representations (e.g. geometric patterns, spectral responses, parametric features) that may be best handled by different network architectures working in concert. For example, Chen \textit{et al.} (2025) reported a CNN-LSTM-A model combining convolutional layers, LSTM networks, and attention mechanisms that achieved a prediction accuracy of 0.993 for the spectral response of all-dielectric trimer metasurfaces exhibiting double Fano resonances \citep{chen_prediction_2025}. We believe that these hybrid network approaches are improving the stability and fidelity of AI-driven design, which can further offer new possibilities for designing complex metamaterials. Additionally, integrating large language models appears promising because, as Kim \textit{et al.} (2025), Zhang \textit{et al.} (2025) and Lu \textit{et al.} (2025) demonstrated, it lets researchers achieve comparable results with less ML expertise and less code \citep{kim_nanophotonic_2025,lu_learning_2025, zhang_chat_2025, zhang_chat_arxiv_2025}.

Another major trend is the integration of physical laws and domain knowledge directly into DL models, a practice known as PINNs. By embedding principles like Maxwell’s equations or partial differential equations (PDEs) into the training process or network architecture, these PINNs can significantly reduce the need for large training datasets and improve model reliability. For instance, a convolution-based PINN with U-Net backbones accurately simulates near-field and far-field responses with speed improvements of up to $10{,}000\times$ over traditional solvers\citep{medvedev_physics-informed_2025}. Ongoing works may seek to extend this framework by incorporating multiple physical domains (electromagnetic, thermal, mechanical) into AI models. This trend is expected to grow as it reduces the need for massive training datasets, but the long training time and high GPU memory consumption represent major obstacles for this method.

Newer and complex AI architectures are showing an ever-growing potentiality. A refined GAN paired with an agent model based on the Swin Transformer \citep{liu_swin_2021} has enabled efficient generation of metasurface patterns from spectral data, achieving MSE as low as $6\times10^{-3}$ between simulated and generated spectra\citep{wang_generative_2025}. Likewise, a U-net with CGAN framework has facilitated forward and inverse design of terahertz metasurfaces with multifunctional responses, achieving inverse prediction accuracies exceeding 94\% \citep{xia_deep-learning-assisted_2025}. While these sophisticated architectures yield performance improvements, it's important to balance these gains against the increased computational costs, extended training times they entail, and a potential greater demand for training data sets. In some cases, a simpler model may offer a more efficient solution for less complex design challenges.

In addition to the supervised methods, unsupervised learning offers an alternative framework for metamaterial design. One approach uses the K-Nearest Neighbor (KNN) algorithm, which requires fewer data and computational resources than NNs. KNN clusters and interpolates metamaterial configurations to yield new geometries with defined property combinations without direct supervision. Recently, Fan \textit{et al.} (2025) reported an inverse design method for optical power splitters that combined KNN with particle swarm optimization \citep{fan_novel_2025}. This unsupervised learning method offers a fresh perspective on the inverse design of photonics. We believe that unsupervised learning models can be further extended to other metamaterial design challenges.

\section{Conclusions}
In this review, we have provided a comprehensive overview of data-driven approaches in nanophotonics, with a particular focus on the design and optimization of AI-enabled metadevices. Our discussion highlighted the significant strides achieved through both image-based and parameter-based DL methods. Advanced techniques including CNNs, RNNs, GANs, VAEs, and most recently, transformers have demonstrated their ability to efficiently navigate complex, high-dimensional design spaces and account for intricate physical phenomena such as mutual coupling effects. These methodologies not only overcome the computational limitations of traditional design approaches but also enable the rapid prediction and inverse design of multifunctional photonic architectures.

To conclude, the fusion of artificial intelligence with nanophotonic engineering marks a transformative shift in meta-devices development. As described throughout this review, AI-driven strategies hold immense promise for enhancing design precision, accelerating optimization processes, and ultimately facilitating the development of next-generation photonic platforms. Looking ahead, future research is expected to delve deeper into hybrid network architectures, physics-informed learning, and attention-based models, thereby broadening the scope and impact of data-driven nanophotonic design.

\paragraph{Conflicts of interest} There are no conflicts to declare.

\paragraph{Acknowledgements} This work was supported by the John L. and Genevieve H. McCain endowed chair professorship at The Pennsylvania State University.

%%% END INSTRUCTIONS %%%

\bibliographystyle{IEEEtranN}   % natbib-compatible, numeric [xx]
\bibliography{references}       % your references.bib file
\end{document}